\documentclass[]{egpubl}
\JournalSubmission    

\usepackage{t1enc,dfadobe}

\usepackage{graphicx}
\usepackage[labelfont=bf,textfont=it]{caption}
\usepackage{amsmath,amssymb}

\usepackage{color}
\definecolor{turquoise}{cmyk}{0.65,0,0.1,0.1}
\definecolor{purple}{rgb}{0.65,0,0.65}
\definecolor{dark_green}{rgb}{0, 0.5, 0}
\definecolor{orange}{rgb}{0.8, 0.2, 0.2}
\definecolor{red}{rgb}{1.0, 0.1, 0.1}

\newcommand{\kxc}[1]{{\color{red}[[Kevin: #1]]}}

\newcommand{\xh}[1]{{\color{dark_green}#1}}

\newcommand{\vfignudge}{\vspace{-0.08in}}
\newcommand{\mypara}[1]{\noindent \textbf{#1.}}

\def \score {\textup{score}}

\def \score {\textup{score}}

\def \user {\textup{user}}

\def \detail {\textup{detail}}

\title[Sketch-to-Design]{Sketch-to-Design: Context-based Part Assembly}
\author[Xie et al.]{Xiaohua Xie \quad Kai Xu \quad Niloy J. Mitra \quad Daniel Cohen-Or  \quad Baoquan Chen}

\begin{document}

\maketitle

\begin{abstract} Designing 3D objects from scratch is difficult, especially when the user intent is fuzzy without
a clear target form. 
In the spirit of modeling-by-example, we facilitate design by providing reference and inspiration from existing model contexts. We rethink model design as navigating through different possible combinations of part assemblies based on a large collection of pre-segmented 3D models.
We propose an interactive sketch-to-design system, where the user sketches prominent features of parts to combine. The sketched strokes are analyzed individually and in context with the other parts to generate relevant shape suggestions via a design gallery interface. As the session progresses and more parts get selected, contextual cues becomes increasingly dominant and the system
quickly converges to a final design. As a key enabler, we use pre-learned part-based contextual information to
 allow the user to quickly explore different combinations of parts.
 Our experiments demonstrate the effectiveness of our approach for efficiently designing new variations from existing shapes.


\end{abstract}

\section{Introduction}
\label{sec:intro}

Conceiving shapes from scratch is difficult since early concepts are often fuzzy, ambiguous, and not fully formed~\cite{Insitu:SIGA11}. In the early design stages, artists typically explore multiple conceptual options, without prescribing their details. For examples, artists prefer to start with rough sketches, which they progressively over-sketch to eventually converge to a conceptual shape. With a similar motivation, the recent ShadowDraw system~\cite{Lee:2011:ShadowDraw} uses a data-driven approach to guide the artists to create better and well-proportioned sketches. The system, however, does not immediately generalize to 3D since the evolving conceptual shape cannot be observed or edited  from multiple view directions. We introduce a sketch-to-design interactive system that instantly converts user sketches to part-based 3D geometry, thus retaining the fluidity of the sketching process, while allowing easy 3D model creation.

\begin{figure*}
 \includegraphics[width=\linewidth]{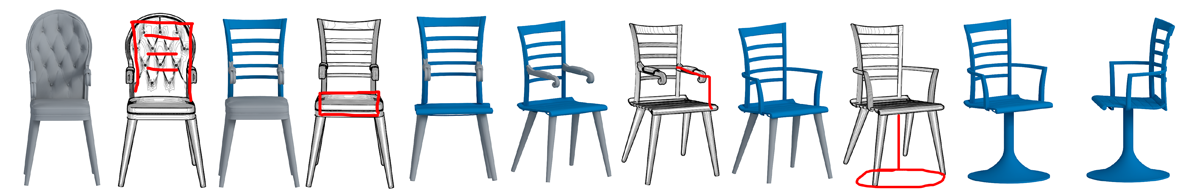}
 \centering
  \caption{Starting from an arbitrary 3D chair model~(left), the user sketches (in red) over a  light ``shadow" of
  the chair. As the session progresses, the user can rotate the current model and sketch over more convenient viewing directions. The
  user strokes along with symmetry and context-information from pre-analyzed database models are used to retrieve, deform, and
  snap parts to provide modeling suggestions to the user. Effectively, the strokes helps guide a part-based design space exploration. }
\label{Fig:Teaser}
\end{figure*}

A successful 3D modeling system should be simple, interactive, intuitive to use, and provide multiple design options for different user preferences. 
In our system, modeling amounts to navigating a space of
\emph{mix-and-match models}, with the user sketches and context information driving the navigation.
The user simply sketches prominent features and desired shapes of parts, while the system computationally retrieves the compatible parts and handles low-level operations to assemble and warp the parts together. As the user progressively explores and selects model parts, fewer model parts with compatible context clues are left to choose from, thus narrowing down design possibilities. As an analogy, think of autocomplete option in textual search engines --- as the design session progresses, modeling speed increases with fewer part options to select from (see Figure~\ref{Fig:Teaser}).

Advances in consistent decomposition of models into parts (e.g.,~\cite{Kalogerakis:2010:LMS,Huang:2011:JSS,Sidi:2011:CS}) motivate part reuse for model creation. We make use of relative placement and context information across parts in large collections of semantically segmented parts to allow the user to intuitively select, position, and glue parts to produce novel models.  Specifically, we analyze a large set of segmented models to learn their contextual relations (e.g., part pairs in contacts, in symmetry, or in similarity of geometry properties) and use the relations for smart design exploration. Thus, we bypass the difficult task of understanding semantics of the parts and their inter-relations.

Inspired by ShadowDraw~\cite{Lee:2011:ShadowDraw}, we present an interactive system where the user roughly sketches parts over a canvas that displays the current 3D model in the background. We continuously analyze the drawing strokes and their context to suggest relevant part combinations to the user via a dynamic gallery, with contextual clues becoming increasing dominant in the later stages of the session. At any stage, the user sketches a part-profile in 2D, while the system suggests multiple part possibilities using their original context. The user selects one such possibility, the part adapts to the design, and the session proceeds (see supplementary video). Note that the user can change viewpoint at any time.

Generally, sketch-based retrieval of parts is difficult. Since parts display less variations compared to whole shapes
the visual cues between different parts are less discriminative than those for whole shapes.
Hence, purely silhouette-based part retrieval is often too ambiguous.
Further, the user's sketch in an open-ended modeling process often tends to be less informative.
%
In contrast to previous attempts (e.g., \cite{Lee:2008:SSA}), we remove such ambiguities by analyzing the contextual relations of the relevant parts.
For example, while designing a chair,  a rounded chair seat may suggest, with higher confidence,
a rounded chair back rather than a square-ish one. The correlated features include geometry properties
internal to the parts (e.g., parallel banisters of a chair back) and not simply their outlines.
During the sketch-to-design process,  we make use of such contextual relations to significantly
reduce the search space of part assemblies, thereby  better assisting the designer to quickly
converge to an appropriate design.
We evaluate the effectiveness of our modeling system using a user study and present
various models generated using our system (see also supplementary video).

\section{Related Work}\label{sect:related_work}

Modeling remains a popular research topic with a host of relevant efforts in recent years. In this section, we focus on a few representative papers relevant to our work.

\mypara{Sketch-based retrieval}
Inspired by Funkhouser et al.~\shortcite{Funkhouser:2003:ASE}, many image-based approaches have been proposed towards
sketch-based shape retrieval (see~\cite{Shao:2011} and references therein). Earlier, Chen et al.~\shortcite{chen03} propose Light Field Descriptor to encode 2D projections of a 3D shape using a combination of contour- and region-based shape descriptors.
This method, however, does not consider the interior feature lines.
Recently, Eitz et al.~\shortcite{eitz2012} propose a bag-of-words based method to encode view-dependent line drawings of the 3D shapes using both silhouette and interior lines. Subsequently, they learn a classifier based on a large set of human created classified sketches. However, for sketch-based part retrieval, we have to deal with imprecise and less discriminative drawings of shape parts. We adapt sketch-based retrieval proposed by Lee et al.~\shortcite{Lee:2008:SSA}, while achieving robustness using contextual information among pre-analyzed parts.


\mypara{Assembly-based modeling}
As model collections grow, modeling by part assembly provides a quick way to synthesize new models from the existing ones.
In a seminal system~\cite{Funkhouser:2004:MBE}, modeling-by-example rely on shape-based search to find desired parts to assemble. The user provides rough a 3D proxy of the required part, which is then used to query the database of shape parts. While the concept is powerful, the interface is cumbersome requiring users to model, position, and manipulate proxies in 3D.
Although subsequently various sketch-based user interfaces have been proposed~\cite{Shin:2007:MCI:1268517.1268530,Lee:2008:SSA,Fisher:2011}, the methods either require sketching proxy geometry in 3D or restrict view manipulations during any session.

Recently,
Kalogerakis et al.~\shortcite{Kalogerakis:2012:ShapeSynthesis} propose a probabilistic model for automatically synthesizing
3D shapes through automatic model synthesis using training data.
Xu et al.~\shortcite{xu2012} design part crossover for 3D model set evolution based on part assembly with contextual information.
Jain et al.~\shortcite{Jain:2012:3DModelRecombination} study {\em interpolation} of man-made
objects through part recombination. Although such methods produce volumes of shape variations, the methods do not
provide the user with fine-grained control necessary to facilitate interactive design.
Shen et al.~\shortcite{shen2012} exploit the use of part assembly in recovering high-level structures from single-view scans
of man-made objects acquired by the Kinect system.

\mypara{Data-driven suggestions}
We draw inspiration from data-driven suggestions for modeling~\cite{Chaudhuri:2010:DDS,Chaudhuri:2011:PAM} and
shadow-guided sketching~\cite{Lee:2011:ShadowDraw}, while sharing motivation from context-based search for models in 3D scenes~\cite{Fisher:2010,Fisher:2011}.
Our focus, however, is to enable an interactive sketch-to-design system to support conceptual design. Thus we continuously present the user with a variety of suggestions, with the user actively guiding the part-based shape space exploration.
The context information, updated on the fly, allows robust retrieval of relevant parts thus allowing the user to sketch imprecisely.

\begin{figure*}[t!]
  \centering
  \includegraphics[width=.9\textwidth]{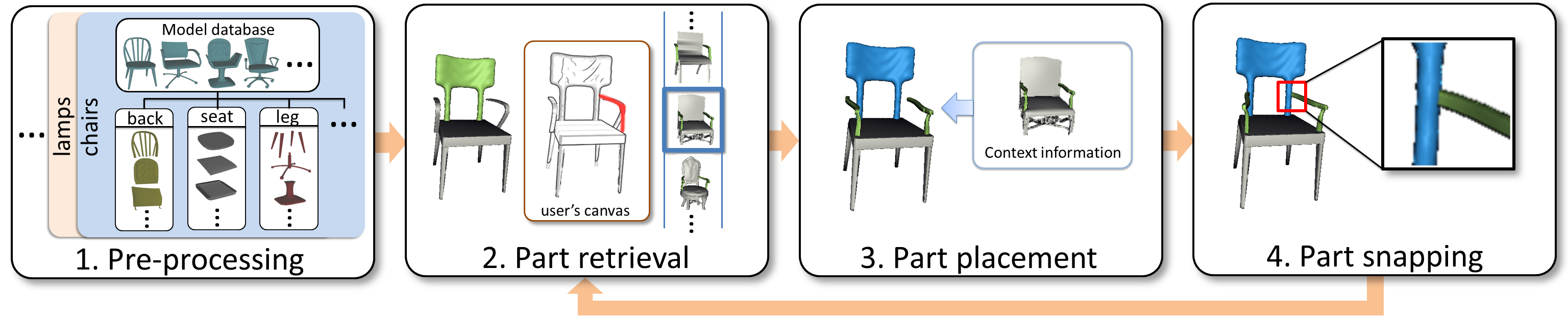}
  \caption{System pipeline.}\label{fig:pipeline}
  \vfignudge
\end{figure*}

\begin{figure}[b!]
  \centering
  \includegraphics[width=0.8\columnwidth]{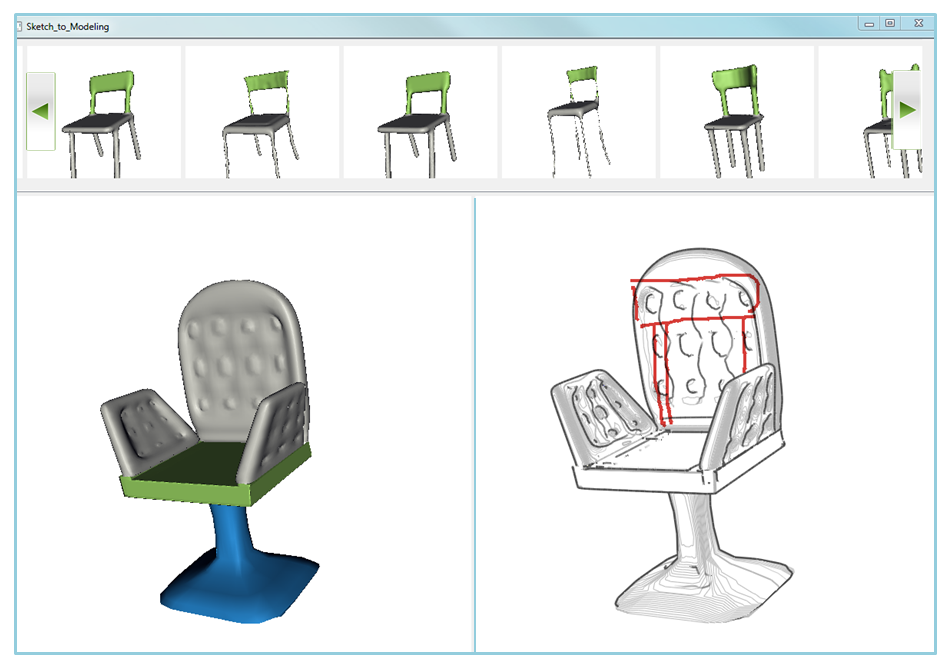}
  \caption{A snapshot of our context-based sketch-driven 3D modeling interface.
The canvas for sketching is on the bottom right panel; the suggestion panel displaying a gallery of relevant parts is at the top; and the panel showing the evolving model is at the bottom left.
}\label{fig:interface}
\vfignudge
\end{figure}

\section{Overview}

Our system comprise of an offline phase (Section~\ref{sec:preproc}) to pre-analyze a 3D candidate part database
and an online interactive modeling system (Section~\ref{subsec:retrieval}) driven by sketch-driven context-based part retrieval and assembly.

\mypara{Offline pre-processing}
We assume the availability of 3D model collections~(e.g., \cite{Ovsjanikov:2011}).
We consider 4 classes of models in our setup: aeroplanes, chairs, lamps, and vases and allow users to explore part-based assemblies for creating model variations inside these classes.
For each class, we compute a representative shape.
Our dataset is pre-segmented and the parts are grouped by their semantic labels
(e.g., legs, slats, seat, wings, handle, etc.) and aligned using upright orientation of the original models.
We then extract contextual information among the parts.
In the modeling session, we use these information to retrieve, place, and connect the parts.

\mypara{User interface}
The modeling interface consists of three parts~(see Figure~\ref{fig:interface}):
(i)~a canvas for sketching the model,
(ii)~a suggestion panel displaying a gallery of relevant parts retrieved from the candidate part database using the context-information, and
(iii)~a panel showing the current design.
The user conveys her design intent via free-hand sketches indicating 2D silhouettes,
or 2D edges indicating prominent geometric features.
At the beginning, a reference model, rendered in an NPR (line drawing) fashion in the canvas, is displayed to the user.
The user can draw strokes over the reference model, in the similar spirit to ShadowDraw.
However, the reference model provides not only a reference for user's drawing but also
the context for part retrieval and placement.
The user can change viewpoint at any point.

\mypara{Modeling}
The user starts by selecting a model types (e.g., chair, vase) as we show the representative model.
Then, the user progressively constructs a complete 3D model in a part by part fashion using a sketch-based interface (see supplementary video and demo).
Modeling proceeds as follows (see Figure~\ref{fig:pipeline}): 

\begin{enumerate}

\item {\em Reference-guided part sketching.} The user, inspired and guided by the related part of the reference model, over-sketches a shape part on the canvas.
The sketch not only provides geometric hints for the part but also about their size, position, etc.

\item {\em Context-based part retrieval.} Based on the user's sketch, we query the candidate part database and
return a sorted list of candidate parts in the descending order of relevance based on degree of 2D-3D matching between the sketch and the candidate part, and also
 contextual information with finalized parts (see Section~\ref{subsec:retrieval}).

\item {\em Context-aware part placement.} From the retrieved candidate list, the user selects a part while our system
automatically computes an appropriate transformation to fit the selected part into the current model.
Again we rely on contextual information for this step (see Section~\ref{subsec:assembly}).

\item {\em Contact-driven part snapping.} To further enhance the quality of the constructed model, we perform a contact-driven part warping
to snap the contact points of the part to the finalized parts (see Section~\ref{subsec:assembly}).
\end{enumerate}

After each part placement, our system automatically suggests a list of adjacent parts to be added next. The user simply selects the one she likes. Effectively, the user strokes are used to only guide selection for part-based modeling (see supplementary video).

\section{Preprocessing}
\label{sec:preproc}

In the preprocessing step, we organize the input database of 3D candidate parts
to support the online parts query for assembly-based modeling.
First, we collect several sets of 3D shapes, each belongs to specific shape classes.
For each class, we compute a representative shape as the center shape in the space of Lighting Field
Descriptor~\cite{chen03}, which acts as representative model for the class.

To build a candidate part-database, we perform consistent segmentation within each class to decompose the models
consistently into different functional/major parts. For example, a chair model is decomposed into four major parts:
back, seat, armrest, and legs. Consistently segmented and labeled datasets can be obtained using
 Kalogerakis et al.~\shortcite{Kalogerakis:2010:LMS}.
For models with multiple components, we use the co-segmentation method of Xu et al.~\shortcite{Xu:2010:SCS}.
Further, when automatic results are unsatisfactory (e.g., vases and lamps), we manually correct the results.
%
After segmentation, all the candidate parts are grouped into semantic category and
aligned with the (manually assigned) common orientation for all the database models within the same class.
The upright orientation is used to compute the initial alignment for the candidate part placement.

In order to support sketch-based part retrieval, we pre-compute the suggestive contours~\cite{DeCarlo:2003:SCC} for each part
from $169$ different positions uniformly sampled on the view sphere. For each such suggestive contour image, we pre-compute features as described in Section~\ref{subsec:retrieval}. To support context-based part assembly,
we pre-analyze each input model to learn the mutual contextual information.
Specifically, for any pair of parts that are adjacent in the original model, we compute the
mutual spatial relations between their oriented bounding boxes (OBB).
Within each model, we detect the global reflectional symmetry as well as the inter-part symmetries~\cite{Mitra:2006:PAS}.
Finally, if a part is self-symmetric and its symmetry reflectional axis is aligned with that of
the global symmetry of the whole shape, we record the part to be self-symmetric.

\section{Augmented Sketch-based Part Retrieval}
\label{subsec:retrieval}


In order to retrieve proper candidate parts using user sketches we use a method similar to Eitz et al.~\shortcite{eitz2012}, which uses a bag-of-words features for sketch-based 3D shape retrieval. Additionally, we also consider contextual information whereby
similarity is measured not only based on the user's sketch, but also taking into account the already placed parts that are adjacent to the current one.
Specifically, we introduce two contextual constraints to ensure the consistency of
both the overall shape and geometric details between the current part and already placed  adjacent parts.


\begin{figure}[t]
  \centering
  \includegraphics[width=0.99\linewidth]{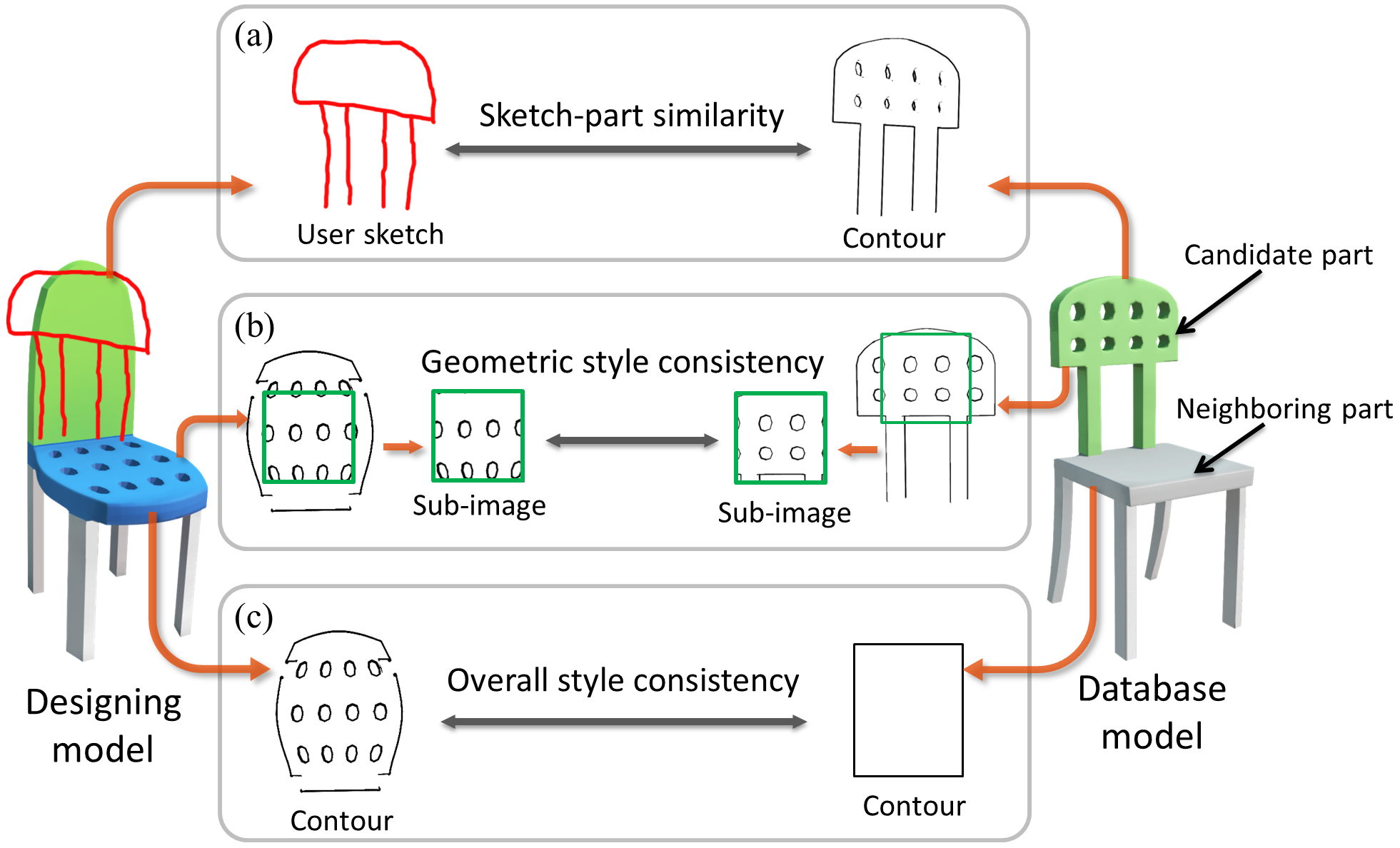}
  \caption{The relevance score for part retrieval contains three components: (a)~the similarity between the user's sketch and
  the contour of candidate parts, (b)~the contextual consistency of geometric style and (c)~the overall style.}
  \label{fig:matching}
  \vfignudge
\end{figure}

\subsection{Relevance score}
Let $c_{\user}$ denote the user's sketch. We measure the similarity between $c_{\user}$ and a candidate part
as a \emph{relevance score} that combines both the sketch-part similarity, i.e.,
the similarity between the projected 2D contours (including both silhouette and interior feature lines) of a part with the user's sketch,
and the part-to-part consistency, which measures the consistency between the already placed neighboring parts.
Thus, the part-part similarity incorporates the contextual information.
Specifically, the relevance score for a candidate part $p \in M$ is defined as
\begin{equation}\label{Eq:rel_score}
 \begin{split}
\score (p) = & s(c_{\user}, c(p)) + \frac{1}{|\Omega|}\sum\limits_{p' \in {\Omega}}(\lambda_1 s_{detail}(c(p'),c(p))\\
 &+\lambda_2 s(c(p'),c(\theta_M(p')))),
 \end{split}
\end{equation}
where $s(\cdot,\cdot)$ measures the similarity between two 2D contours,
emphasizing mainly the large scale line features such as silhouettes.
In particular, the similarity measure $s_{\detail}(\cdot,\cdot)$ is
confined within the silhouettes and focuses only on the interior geometric details.
This is achieved by taking a small window at the center of the bounding box of the 2D contours
and measuring the similarity of the contours within that window.
The window size is set as the $2/3$ area of the (normalized) bounding box
with $\Omega$ denoting the set of adjacent parts $p$ which are already placed.
The corresponding part in model $M$ that shares the same category with $p$ is denoted by $\theta_M(p)$.

\begin{figure}[t!]
  \centering
  \includegraphics[width=0.7\linewidth]{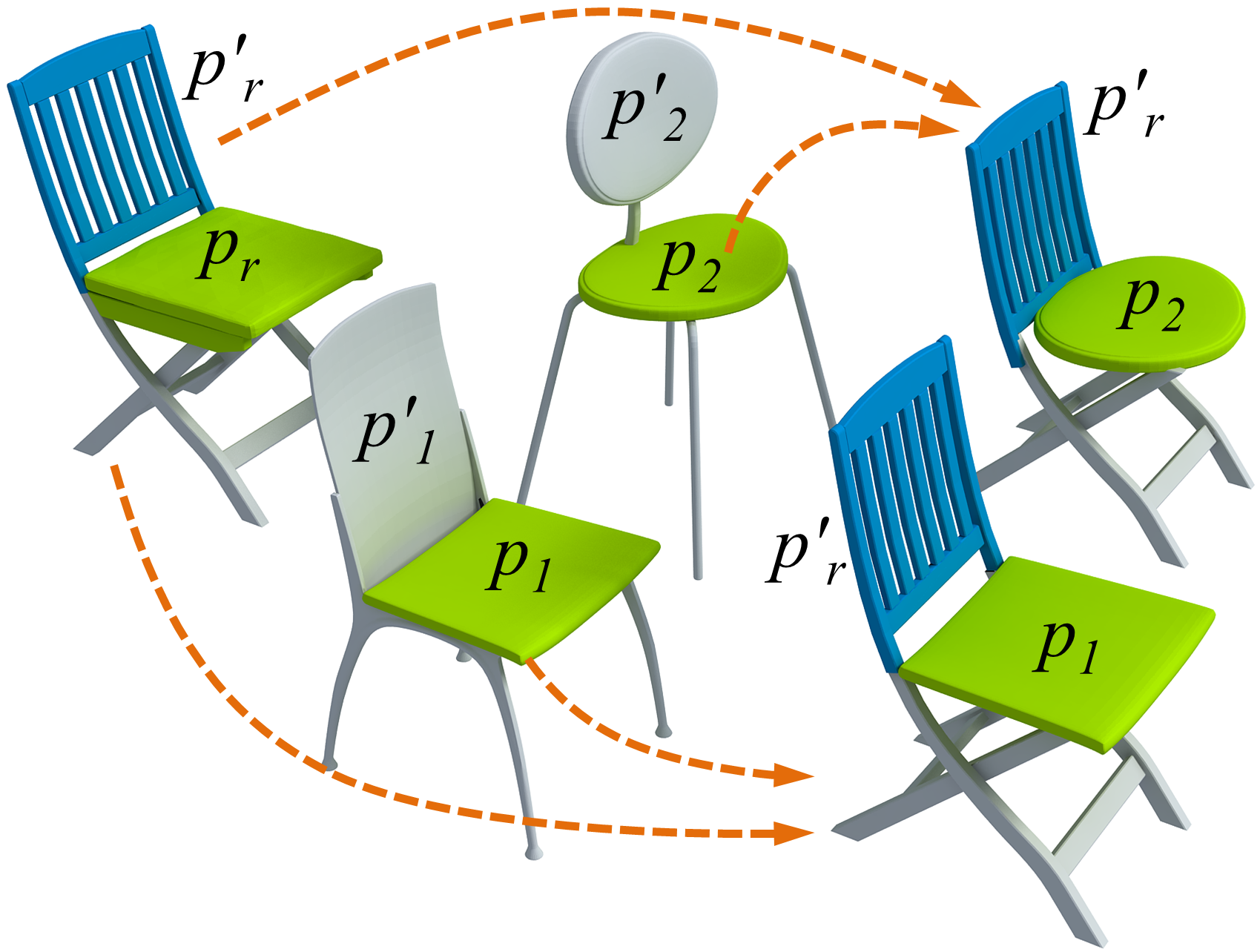}
  \caption{Consider parts $p_1$ and $p_2$ from two candidate models to replace part $p_r$. Taking into account the context and their fitness function with $p_r'$, part $p_1$ fits better as it is consistent with the 2nd term in Equation~\ref{Eq:rel_score}. Note that multiple retrieved parts along with their fitness scores are presented as suggestions to the user.  }\label{Fig:connect_harmonicity}
  \vfignudge
\end{figure}

The first part of Equation~\ref{Eq:rel_score} measures the similarity between the user's sketch and the
the contour of candidate parts (Figure~\ref{fig:matching}a).
The second part accounts for the contextual information,
where the first term $s_{\detail}(c(p'),c(p))$ focuses on the consistency of geometric style between two parts,
indicating that parts with similar geometric texture match better (Figure~\ref{fig:matching}b).
The second term measures the consistency of the overall shape style between two parts (Figure~\ref{fig:matching}c).
For example, a squarish back of a chair matches better with a squarish seat than a roundish one (see Figure~\ref{Fig:connect_harmonicity}).
The weights $\lambda_1$ and $\lambda_2$ are used to tune the importance of the two contextual constraints.

\subsection{Feature representation.}
In order to retrieve a 3D part according to the 2D sketch $c_{\user}$, we measure the similarity between
$c_{\user}$ and the suggestive contours $c(p)$ of a part $p$ obtained from the user's current viewpoint.
Enforcing contexual consistensy requires the comparison between two 3D parts, but the matching is view-dependent. However,
as we are only concerned with the comparison between parts in the same category, we can compute a common view for the parts
and measure the similairty between their suggestive contours from that common view.
Specifically, for each part category, the common view is computed as the direction along the
shortest PCA axis of the averaged OBB of all parts in that category.
Since all the parts in the same category are aligned, we compute the averaged OBB.
The final similarity between two parts is the average of the contour similarities measured from both orientations along the common view.
Thus, both sketch-part and part-part matching reduces to a image matching problem.

Both the 2D contours and the user's sketches are treated as 2D images for which the feature representation is based upon a
bag-of-words (BOW)~\cite{Sivic:2003:BOW} model. In our system, we scale the images being matched into $320\times 320$ pixels.
For each image, we generate $32 \times 32=1024$ key points evenly distributed over the image by sampling on a regular grid and extract  local features around each key points.

We adopt the Gabor local line-based feature~(GALF)
along with the optimal parameters suggested by Eitz et al.~\cite{eitz2012}.
Specifically, 4 orientational filters are all used to compute the Gabor response for $4 \times 4=16$ cells around each key point.
For each orientation, its average response within a cell is used to construct the final features for that cell.
Thus, each feature vector has a size of $4 \times 4 \times 4 =64$ per key point,
and $1024$ feature vectors per image. Before extracting the features, we apply a skeletonization algorithm~\shortcite{Zhang:1984:FPA:357994.358023} to attain a unified line width for both the user's sketch and contours.

Based on the features extracted from the contours of all candidate parts and views, we build a "visual vocabulary'' $V=\{w_i\}^j$ by clustering the features, where each cluster centroid is a "visual word". In our experiment, we set the size of the vocabulary as $2500$. Thus, each image in that view is represented by a histogram of occurrences of these visual words $V$.

Finally, we use Term Frequency-Inverse Document Frequency~(TF-IDF) weight~\cite{Witten:1999:MG} to unify the computed histograms.
The TF-IDF balances the occurrence frequencies of visual words in a spacial image and training set by representing $h_i:=(h_i/\Sigma_jh_j)log(N/N_i)$ where $N_i$ and $N$ are the occurrence number
of the visual word $w_i$ and the total number of visual words in the whole training image set, respectively. The visual word occurrence histograms $\{h_i\}_i$ are matched using $\chi^2$ distance, i.e.,
\begin{equation}\label{Eq:ch2}
s(h_i,h_j)=<h_i,h_j>/||h_i||||h_j||.
\end{equation}




\subsection{Part retrieval.}
\label{subsec:part_retrieval}
For online part retrieval, a straightforward approach is to compute the relevance score using Equation~\ref{Eq:rel_score}
for each candidate part in the database
and then obtain a list of most relevant candidates with the maximal relevance scores.
This, however, gets expensive for a large-scale database.
Instead, we employ the inverted index structure~\cite{Witten:1999:MG} to reduce the search space.

Specifically, we learn three visual vocabularies $V_k, k=1,2,3$ for the three terms in Equation~\ref{Eq:rel_score}, one for each term. Based on the visual vocabulary $V_k$, for each query, we build an index $Ind_k$ to the subset $A_k$ of the database that contains the images
sharing visual words with the query. Consequently, the final search space for that query is formed by the intersection of corresponding three image subsets, i.e., $E:=\cap^3_{k=1} {A_k}$. Because the histogram of visual words $\{ h_i \}_i$ is always sparse, the number of images in $E$ is much smaller than that in the original database, the search time can be greatly reduced.

\subsection{Suggesting adjacent parts.}
Once a part $p$ is placed, we suggest candidates for its adjacent parts $p'$ yet to be placed: We simply take the adjacent parts of the top $K$ parts returned for $p$.
Generally, these suggested parts may contain redundancy. To remove the redundancy, we first perform a k-means clustering over the suggested parts based on the shape distribution
descriptor~\cite{Osada:2002:SD}.
Then, we show only the parts nearest to the centers of clusters.
 Like the retrieved parts, the suggested parts are also displayed in the suggestion panel to inspire the user to proceed with modeling.
The user can either pick a part from the suggestions, or ignore the suggestions and sketch instead.

\begin{figure}[t!]
  \centering
  \includegraphics[width=0.9\linewidth]{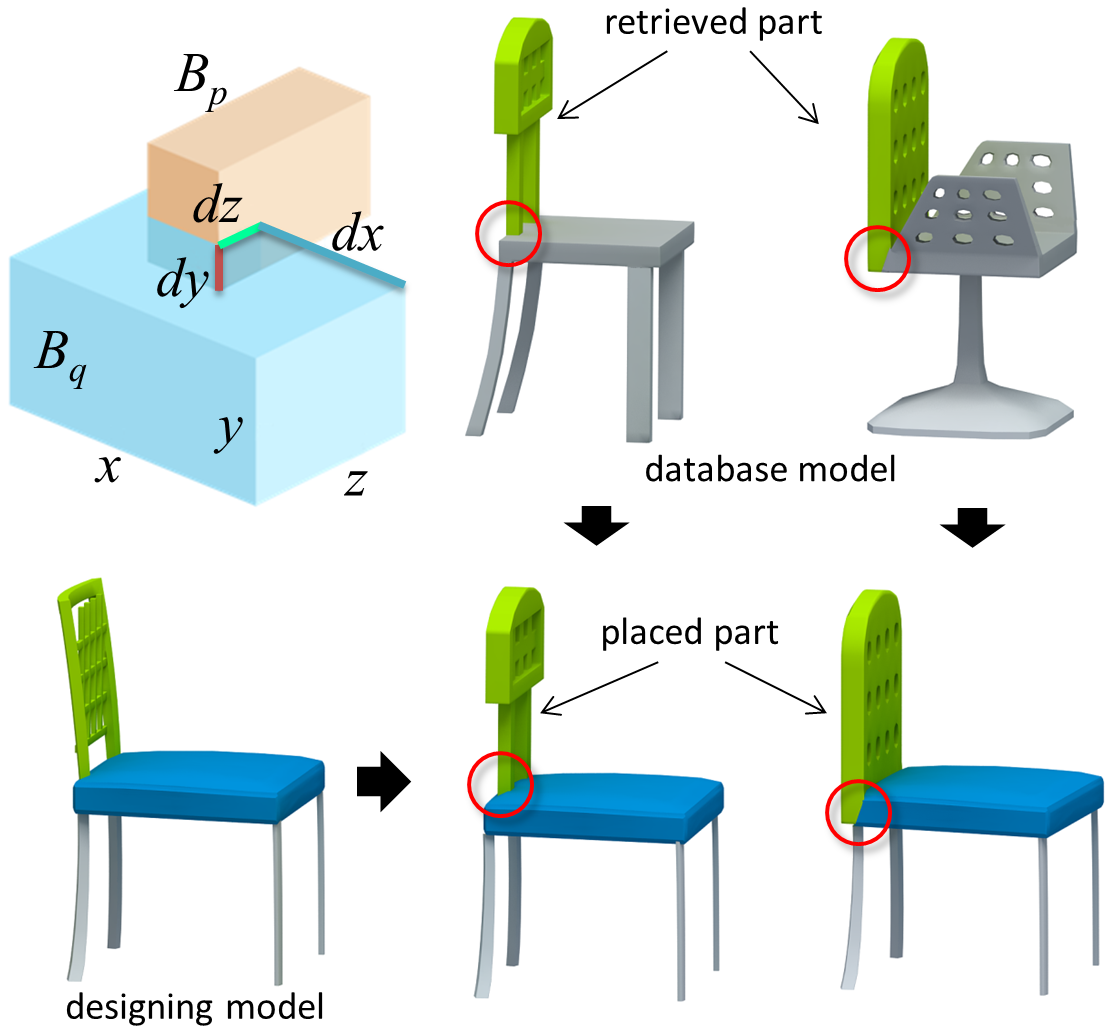}
  \caption{The context-based placement for parts. Top-left figure illustrates ``insertion ratio" between bounding boxes with bottom-left showing the target chair. Middle and right columns show the source chairs with the back (in green) and their placements (bottom) that respect their original ratio with the seat (in blue).
}\label{Fig:contextual_placement}
\end{figure}

\section{Part Assembly}
\label{subsec:assembly}

Once the user selects a part $p$ from the candidate list, we automatically fit $p$ to the user's
sketch $c_{\user}$ through adjusting the size of the 2D bounding box of $p$ projected from the current view.
After that, $p$ is fitted into the target model being built through a context-aware part placement step
and a contact-driven part snapping step.

\subsection{Context-aware part placement}
Suppose that $p$ and $q$ are two adjacent parts in the source model,
$q'$ is the already placed counterpart of $q$ in the target model, meaning that $q'$ and $q$ share the same semantic category.
Our goal is to connect $p$ onto $q'$ reasonably.
To achieve a reasonable placement, we define a set of placement rules $\textbf{R1}$ $\sim$ $\textbf{R3}$,
which are based on prior-knowledge and the contextual information pre-analyzed from the source models:
%
%
\begin{enumerate}

\item \textbf{R1: Insertion ratio preservation.}
       Suppose $B_p$, $B_q$, and $B_{q'}$ are the OBB's of part $p$, $q$, and $q'$, respectively.
       When placing $p$, we maintain the \emph{insertion ratios} of $B_p$ over $B_{q'}$ in the model
       being built with respect to that of $B_p$ over $B_q$ in the source model.
       Given two neighboring OBBs $B_p$ and $B_q$, we measure the insertion ratios of $B_p$ over $B_q$ as
       $dx/x$, $dy/y$, and $dz/z$, where $dx$, $dy$, and $dz$ are the penetration amount of $B_p$ over $B_q$
        (see Figure~\ref{Fig:contextual_placement}).
       By preserving the insertion ratios, the parts can be placed in a same relative position as in the source model.
\item \textbf{R2: Center alignment.}
       Some neighboring parts (e.g., the back and seat of a chair) are both self-symmetric and their reflectional axes
       are aligned with each other in the source model.
       The pre-analyzed constraints are
       applied during the part placement (if applicable), simply by re-aligning the two parts through aligning
       their reflectional axes.
\item \textbf{R3: Inter-part symmetry preservation.}
       Inter-parts symmetries (e.g., the two armrests of a chair) are also pre-detected. Thus, once a part is placed,
       its symmetric counterpart is retreived from the source model and automatically placed according to the symmetry.
\end{enumerate}


\subsection{Contact-driven part snapping}
\begin{figure}[b!]
  \centering
  \includegraphics[width=\linewidth]{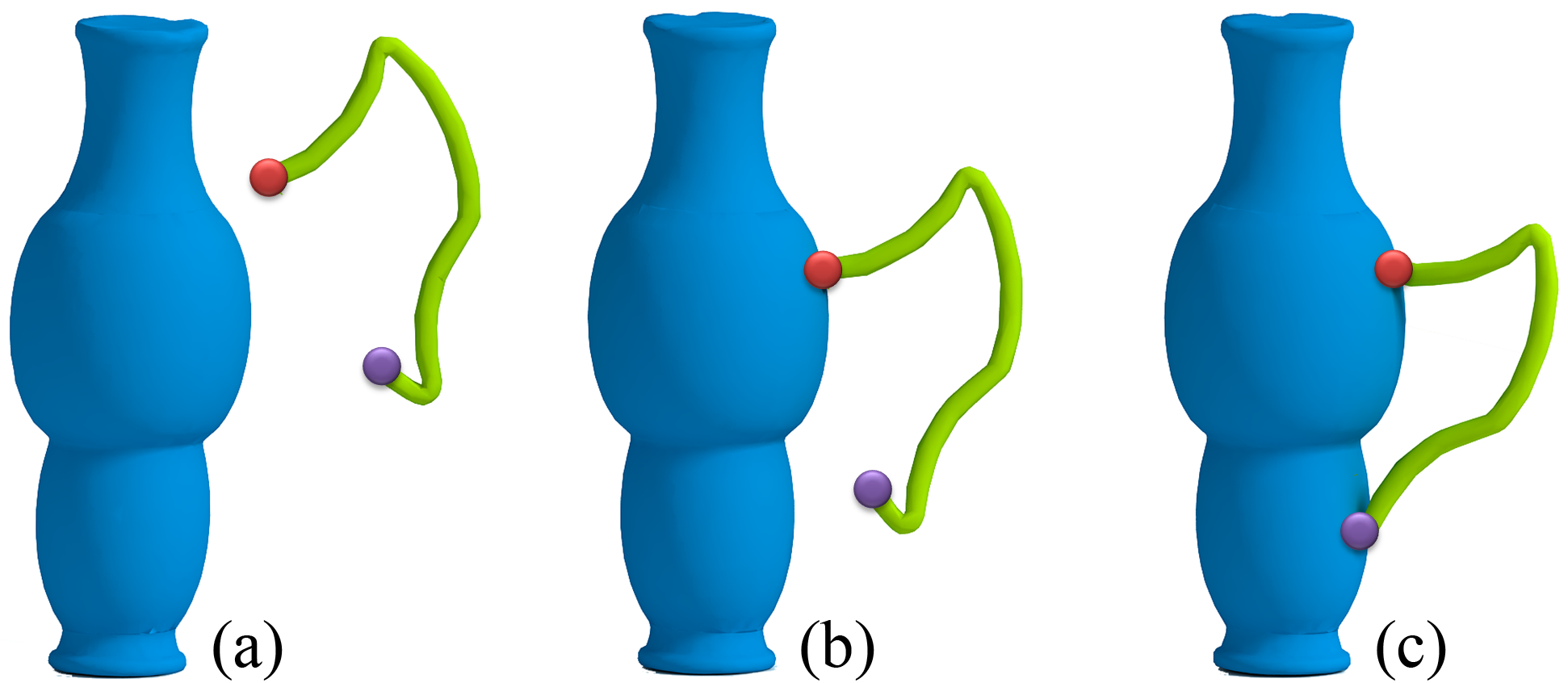}
  \caption{Snapping overview: (a)~Initial placement of a handle with two connecting points;
  (b)~the handle is deformed to snap to the vase body by first snapping the red contact point;
  and (c)~then the purple contact point.
}\label{Fig:docker_snapping}
\end{figure}

After the part placement step, neighboring parts are well connected for most of the cases.
However, there may still be parts are not well connected due to the discrepancy of part size and/or geometry,
where the parts may need (non-rigid) deformation to achieve a better snapping.
We address this using a contact-driven part snapping.
During the offline pre-segmentation, we have recorded the connection points between any two neighboring parts. These connection points
are used as contact points to drive the parts to deform and snap to their neighboring parts in the target model.
Specifically, after a part is placed, our system will ``drag'' the contact points of the part to the nearest points (or the contact points if existing) of its neighboring parts.
Accordingly, the part is deformed using the shape matching based deformation~\cite{muller05} (see Figure~\ref{Fig:docker_snapping}).

\subsection{Part stitching}
\begin{figure}[t!]
  \centering
  \includegraphics[width=0.9\linewidth]{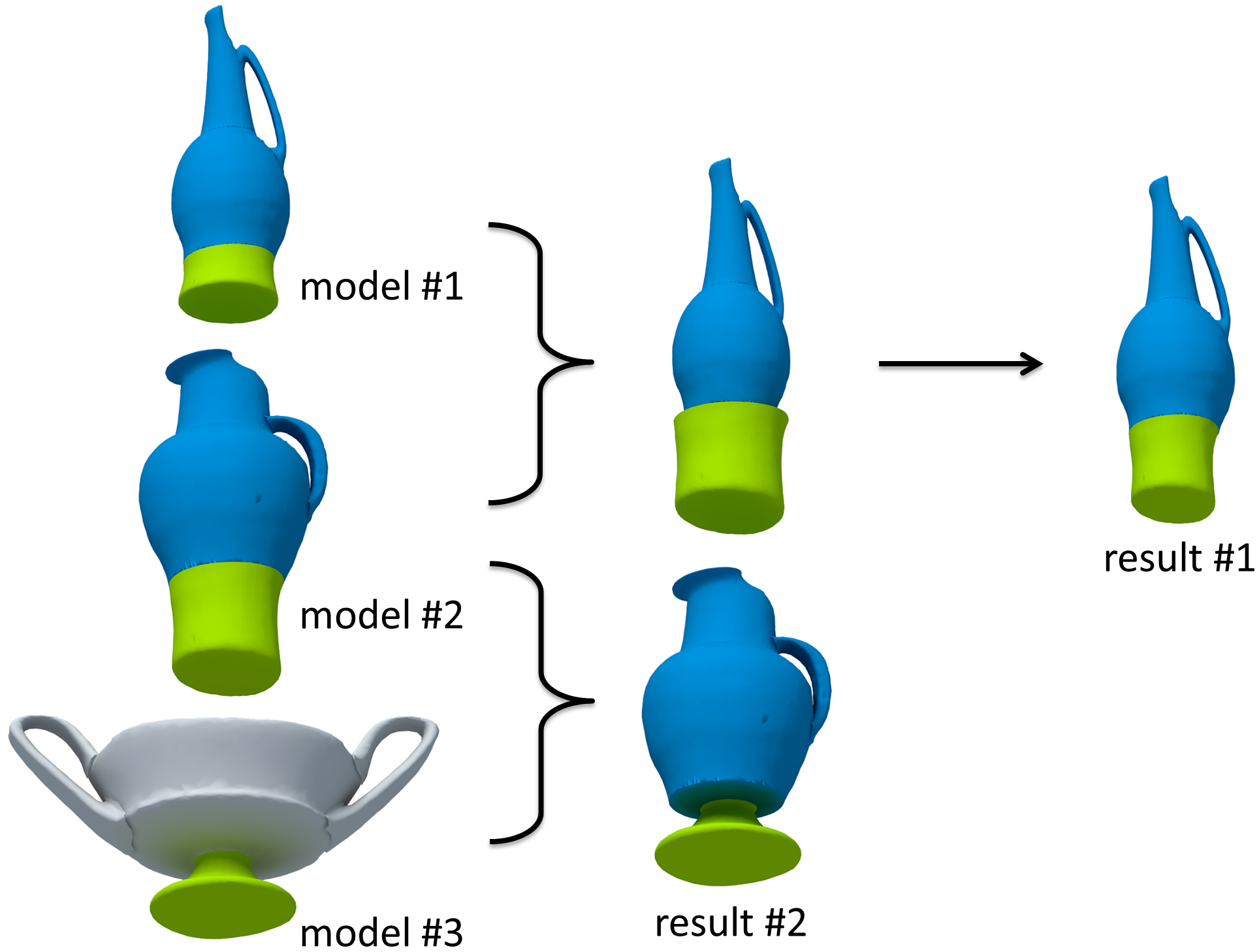}
  \caption{Illustration of local smoothness.
  In the top two models~(left), the transitions between the body and base are both detected as smooth,
  therefore in the result \#1 such smoothness is maintained by scaling the two parts being connected.
  In contrast, the connecting parts in result \#2 are left unchanged.}\label{Fig:local_smoothness}
   \vfignudge
\end{figure}

Finally, we perform part stitching to guarantee quality resulting models.
For two parts to be connected, if the connectors in their source models are both detected as locally smooth,
a scaling is applied to maintain such a local smoothness in the target model. Otherwise, we simply place them together.
After stitching, we sow the mesh along the cutting seam using a mesh stitching~\shortcite{sharf2006}.
The local smoothness are determined by comparing the sizes of bounding boxes of connectors
in two parts (see Figure~\ref{Fig:local_smoothness}).

\section{Result}

We collected a database consisting of 448 3D objects across six categories for our system.
These objects were divided into four subsets, where each contained semantically similar models.
The four subsets were:
chairs and tables ($308$ models),
aeroplanes and birds ($48$ models),
vases ($34$ models), and
lamps ($58$ models).
%
We allow the creation of interesting variations by  mixing the objects from different categories (sharing similar semantic labels). For example, a tabletop can sometimes be selected for the seat of a chair; or, the wings of
a bird can be plugged onto an aeroplane's body.
%
%
In this section, we first evaluate the part retrieval aspect of our method and then the effectiveness of the system via a user study.
A light weight demo system is submitted as accompanying material.

\subsection{Context-based part retrieval}


Context-based part retrieval depends on two
criteria (see Equation~\ref{Eq:rel_score}):
(i)~consistency of geometric style between the retrieved part and the already-placed adjacent ones, and
(ii)~the consistency of the overall shape style between the parts.
To evaluate the effect of the contextual information, we test the part retrieval of our system
under different parameter settings of $\lambda_1$ and $\lambda_2$ in Equation~\ref{Eq:rel_score},
through user evaluation.

With each database, we asked $8$ users to design new models using our system.
For each user, we randomly select from the design sessions $3$ retrieval scenarios
each of which has at least $3$ already placed parts serving as context.
For each scenario, we let another $8$ participants to vote
for the candidate parts (with the same category as the current one) from all the other models in the database
on whether the candidate fits well with respect to the already placed parts, serving as a ``ground truth''.
Thus for each scenario, we obtain a \emph{consistent list} through selecting the top ten candidate parts
based on the positive votes.
If $3+$ of the top $10$ retrieved parts in a scenario overlaps with its consistent list,
we record it as accurate.
The accuracy rate is computed with respect to all the scenarios selected for the database.

\begin{figure}[t!]
  \centering
  \includegraphics[width=.99\linewidth]{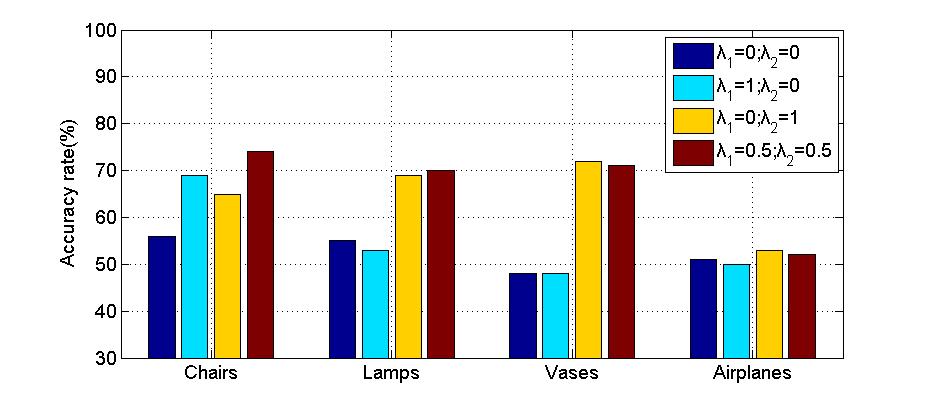}
  \caption{Accuracy rate of part retrieval for four databases under different parameter setting for $\lambda_1$ and $\lambda_2$.}
  \label{Fig:Retrieval_PR}
  \vfignudge
\end{figure}

Figure~\ref{Fig:Retrieval_PR} shows the accuracy rate for different database
under different parameter settings of $\lambda_1$ and $\lambda_2$.
The chair database gains the most in accuracy from the two contextual constraints. This is because
chairs possess prominent geometric styles (in terms of both geometric details and overall shapes)
making the contextual style consistency important.
For lamps and vases, the accuracy gain is dominated by overall shape styles due to the lack of
geometric details. For the airplanes, the contextual information plays a negligible role
possibly because of negligible shape variations in the database.

\begin{figure}[t!]
  \centering
  \includegraphics[width=.99\linewidth]{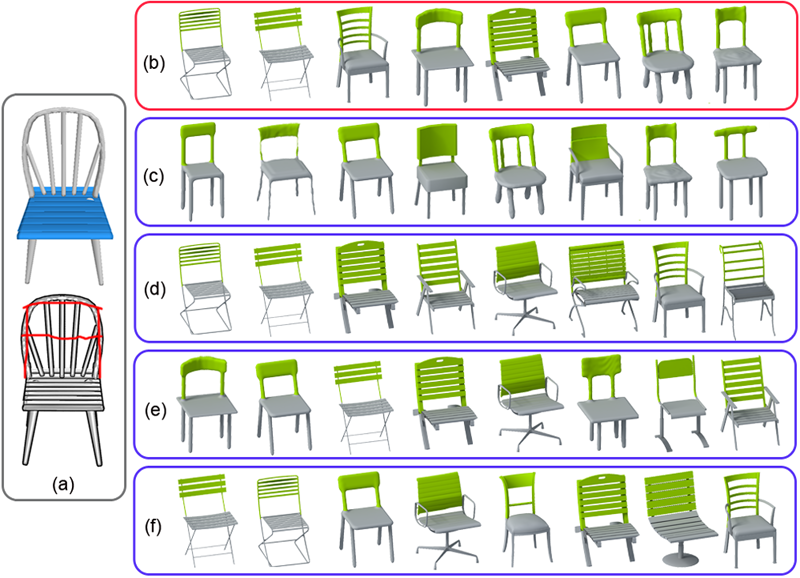}
  \caption{Comparison of retrieved results with and without contextual information:
  (a)~the reference model and its contours with user's sketch, where the seat of chair has been fixed;
  (b)~the ``ground truth'' matching parts voted by $8$ participants;
  (c)~the retrieved results according to only the user's sketch, i.e., $\lambda_1=\lambda_2=0$;
  (d)~the results with $\lambda_1=1.0, \lambda_2=0.0$;
  (e)~the results with $\lambda_1=0.0, \lambda_2=1.0$.
  (f)~the results by considering both the user's sketch and the full contextual information, i.e., $\lambda_1=0.5, \lambda_2=0.5$.
  }
  \label{fig:retrieval_space}
  \vfignudge
\end{figure}

In Figure~\ref{fig:retrieval_space}, we show the effect of the contextual constraints
where we show the retrieval results for a back part of a chair model with the seat fixed under different parameter settings.
When both constrained are disabled ($\lambda_1=\lambda_2=0$),
the retrieval results are affected only by the user's sketches.
However, when they are both enabled ($\lambda_1=\lambda_2=0.5$),
the retrieved parts are more consistent with respect to the neighboring parts and conform better to
the user voted ``ground truth''.

\subsection{User experience}
We test the effectiveness of our interactive modeling system via a user study involving 16 participants.
The group of users consisted of 3D modelers/artists, graduate students in graphics, and non-graphics students (novice users), in roughly equal proportions.
Before the actual testing, participants were allowed to get familiar with the system under the guidance of our developers,  typically <15 minutes.
We conducted two types of user studies: The first one is {\emph{goal-directed modeling}} to test the effectiveness of our
system in returning relevant parts in order to construct a model similar to the goal.
The second one is {\emph{free modeling}} where the user is allowed to freely design
new models by exploring various part assemblies offered by our system.
This tests the ability of our system in supporting the conceptual design of new 3D model.
In all studies we used $\lambda_1=\lambda_2=0.5$.

\begin{figure*}[t!]
  \centering
  \includegraphics[width=0.96\linewidth]{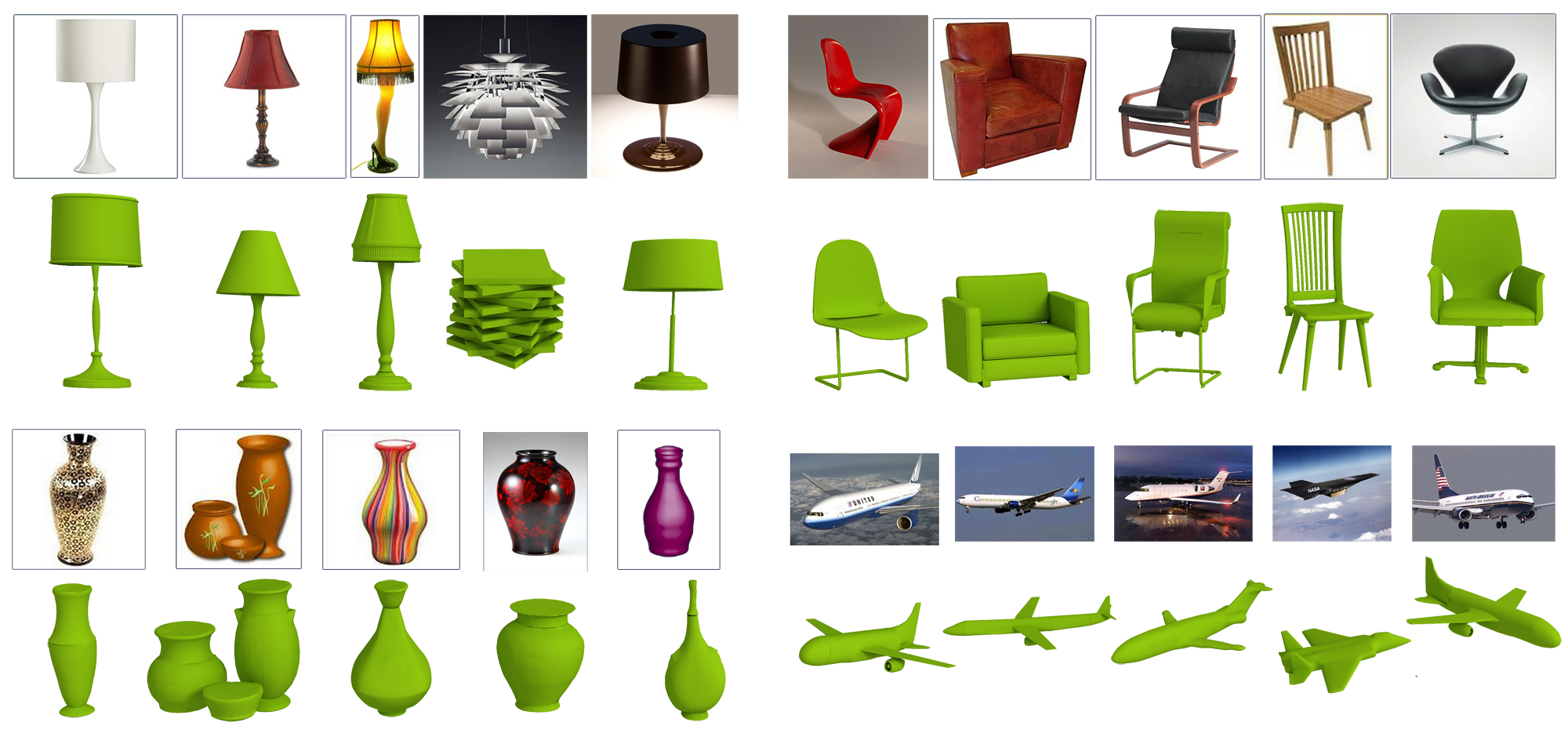}
  \caption{Examples of the ``Google challenge" test. The reference photos are the top five search results from Google engine with the key words ``chair", ``lamp", ``vase", and ``plane", respectively. The objects in each image were modeled by the users using our system. The best result corresponding to each photo is displayed here. }\label{Fig:Google_challenge}
\end{figure*}

\mypara{Goal-directed modeling}
Although our system is designed for open-ended modeling, in order to evaluate the
performance of our retrieval module, we first conduct a goal directed modeling experiment.
We give the users a collection of photographs containing the target object and ask them to build 3D models as similar to the targets as she/he can do.
We have conducted a ``Google challenge'': we used four key words ``chair'', ``lamp'', ``vase'', and ``plane'' to
search for four categories of photos from the Google Image search engine.
For each category, the top five returned images were presented to the used as the goals for modeling.
The modeling results were cross-rated  among the participants. The top modeled shapes (according to the user scores)
for each goal photo are in Figure~\ref{Fig:Google_challenge}.
Additional user study results can be found in the accompanying material.

\begin{table}[t!]
\begin{center}
\begin{tabular}{|c||c|c|}

\hline
 objects & system parameters & average clicks\\
\hline \hline

 chair & $\lambda_1=0,\lambda_2=0$ & 15\\
 & $\lambda_1=0.5,\lambda_2=0.5$ & 9.5\\
 \hline

  & $\lambda_1=0,\lambda_2=0$ & 8\\
table & $\lambda_1=0.5,\lambda_2=0.5$  &6.5\\
 \hline

  & $\lambda_1=0,\lambda_2=0$ & 13\\
airplane & $\lambda_1=0.5,\lambda_2=0.5$  & 11\\
 \hline

   & $\lambda_1=0,\lambda_2=0$ & 10\\
lamp & $\lambda_1=0.5,\lambda_2=0.5$  &8\\
 \hline

   & $\lambda_1=0,\lambda_2=0$ & 13\\
vase & $\lambda_1=0.5,\lambda_2=0.5$  & 10\\
\hline
\end{tabular}
\end{center}
\caption{The average numbers of the temporarily selected models in the database for designing a new model under different retrieval strategies, where $\lambda_1=0,\lambda_2=0$ means no contextual information is considered for the retrieval. }\label{Table:User_Study}
\end{table}

In order to investigate the effect of the contextual part retrieval in the goal-directed
modeling sessions, we record in Table~\ref{Table:User_Study} how many collected models were temporarily selected
(number of mouse clicks by the user) during designing a new model.
Two different parameter settings in Equation~\ref{Eq:rel_score} are compared.
As expected, modeling time is shortened by considering contextual information.

\begin{table}[b!]
  \centering
  \begin{tabular}{@{}|l|r|r|r|r|r|r|r@{}}
    \hline
    {\footnotesize Dataset} & {\footnotesize Size} & {\footnotesize Retrieval} & {\footnotesize Assembly }\\
    \hline
    {\footnotesize Chair} & {\footnotesize 172} &  {\footnotesize 14} &  {\footnotesize 41} \\
    {\footnotesize Table} & {\footnotesize 136} &  {\footnotesize 13} &  {\footnotesize 14} \\
    {\footnotesize Chair$+$Table} & {\footnotesize 308} &  {\footnotesize 14} &  {\footnotesize 20} \\
    {\footnotesize Lamp} & {\footnotesize 58} &  {\footnotesize 13} &  {\footnotesize 21} \\
    {\footnotesize Airplane+Bird} & {\footnotesize 48} &  {\footnotesize 297} &  {\footnotesize 23} \\
    {\footnotesize Vase} & {\footnotesize 34} &  {\footnotesize 294} &  {\footnotesize 77} \\
    \hline
  \end{tabular}
  \caption{Response time of retrieval and assembly in milliseconds on various datasets.}
  \label{tab:Timing}
\end{table}

\begin{figure*}[t!]
\vfignudge
  \centering
  \includegraphics[width=0.9\linewidth]{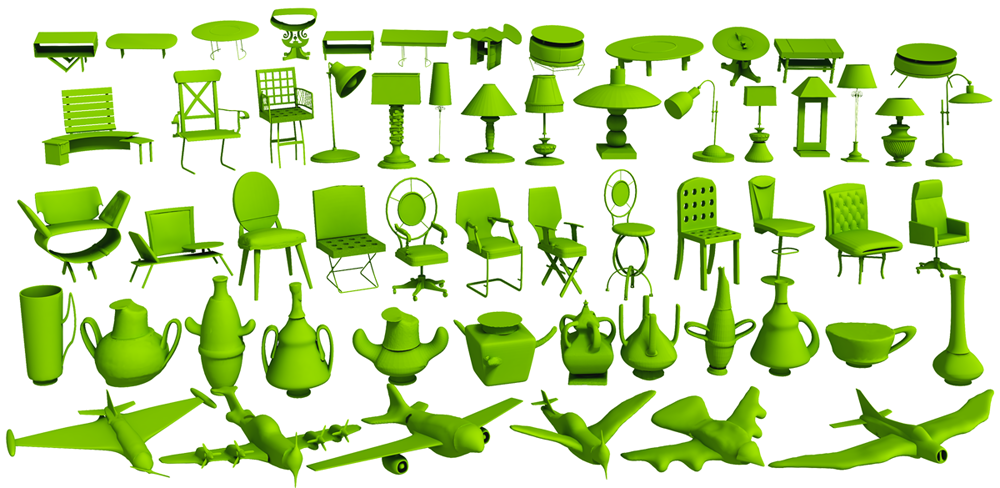}
  \caption{Selection of models created by users of our system.}\label{Fig:user_study_2}
  \vfignudge
\end{figure*}

\textbf{Free modeling.}
In the second user study, we asked the 16 participants to freely create ten different objects using our system.
Here the users are not provided any specific target as goal, except knowing the category she wants to model.
Figure~\ref{Fig:user_study_2} shows a portion of the modeling results produced by the users.
We note that the newly generated models contain a fair amount of variation from the original database models
(see the accompanying material for all the database models).
According to their modeling experience, about $85\%$ of the participants confirmed in questionnaire that they have significantly
benefitted from the intermediate modeling suggestions and the ability to change viewpoint during modeling.
Among the rest $15\%$, most preferred sketching all parts by hand than adopting the automatically suggested ones.

Table~\ref{tab:Timing} shows the response time of retrieval and assembly per part.
Since we use the inverted index structure, the running time on part retrieval does not dependent on the scale
of a dataset, but rather mainly on the amount of geometric variations of the 3D models within the dataset.
The less variations in the model set, the more shared visual words in their BOW features, which
consequently means the search space resulted by the inverted indexing is denser (See Section~\ref{subsec:part_retrieval}).


\textbf{Limitations.} The current system, however, is limited in its creation of finer level of geometric textures (e.g., surface patterns) and micro-structure assembly. Although such textures can possibly be suggested through context (e.g., a beamed chair seat is likely to have a matching beamed back), current sketching tools do not support their direct creation. Also, we focus mainly on conceptual design and produce approximate part assemblies via part snapping, which is guided by only a small number of contact handles.

\section{Conclusion}

We  presented an interactive sketch-to-design system where user provided 2D strokes are used to enable data-driven navigation of design spaces comprising of part-based object variations. For each sketched part, the system suggests plausible 3D shape candidates based on curved-based matching as well as contextual cues, which are extracted from model collections (e.g., models from Google Warehouse).
Inspired by recent success of assisted sketching systems like ShadowDraw, we support 3D modeling that continuously provides design suggestions to the users based on the user strokes. The user can freely change viewpoints during the design session. Retrieved parts are deformed, positioned, and connected to the existing model, again based on context information, as the user implicitly is guided through the possible design space.
We demonstrated the effectiveness of the system in both creative design (i.e., no preset design targets) and modeling from photo inpirations (e.g., re-model from Google Photos) via various user sessions.


With the growing accessibility of model collections and tools to automatically analyze, explore, and handle such collections, we expect to see many data-driven modeling systems. An important direction to explore is to relate geometry to high level object semantics and also bring in support for low-level texture and micro-structural analysis. Ideally, the rich knowledge learned from the database can serve as the ``mind's eye'' for the design system.
A true creative design should allow us to go beyond conventional forms and semantics, to achieve a new level of aesthetics and comfort. For example, a conceptual chair may simply be created out of a few wires, disrespecting the usual functional or semantical decompositions. In the future, we want to bridge this gap in an effort to better support creative design in the early stages of conceptual design.

\bibliographystyle{eg-alpha}
\bibliography{Sketch_3dmodeling}

\end{document}